\documentstyle[preprint,aps]{revtex}

\def\dslash {\partial\!\!\!/}

\begin{document}
%
\title{\LARGE{Structural Aspects of Two-Dimensional Anomalous Gauge
Theories}}
\author{ \sc{ C. G. Carvalhaes$^{\ast\,\ddagger}$, L. V. 
Belvedere$^\ast$, R. L. P. G. do Amaral$^\ast$ },\\
\sc{N. A. Lemos$^\ast$}\\
$\ast$\small{\it Instituto de F\'{\i}sica - Universidade Federal Fluminense}\\
\small{\it Av. Litor\^anea, S/N, Boa Viagem, Niter\'oi, CEP.24210-340, Rio de Janeiro - 
Brasil}
\\
$\ddagger$\small{\it Instituto de Matem\'atica e Estat\'{\i}stica - Universidade 
do Estado do Rio de Janeiro}\\
\small{\it Rua S\~ao Francisco Xavier, 524, Maracan\~a,
CEP. 20559 900, Rio de Janeiro, Brasil}
}
\date{\today}

\maketitle

\begin{abstract}
{\small 
A foundational investigation of the basic structural 
properties of two-dimensional anomalous
gauge theories is performed. The Hilbert 
space is constructed as the
representation of the intrinsic local field algebra generated by the
fundamental set of field operators whose Wightman functions define
the model. We examine the effect of the use of a redundant field
algebra in deriving basic properties of the models and show that
different results may arise, as regards the physical properties of
the generalized chiral model, in restricting or not the
Hilbert space as representation of the intrinsic local field algebra.
The question referring to consider the vector Schwinger model as a 
limit of the generalized anomalous model 
is also discussed. We show that this limit can only be
consistently defined for a field subalgebra of the generalized
model.}
\end{abstract}

\newpage
 
\section{Introduction}

In the last few years, an impressive effort has been made by many physicists
to understand the underlying physical properties of quantum field theory
in two dimensions \cite{AAR91}, as well as to try to picture these models as
theoretical laboratories to obtain insight into more
realistic four-dimensional field theories, and, more recently, to
apply them to low-dimensional condensed 
matter systems \cite{Fradkin91}.  After over a 
quarter of a century of investigations on two-dimensional field theories we 
have
learned that two-dimensional models have also the value of providing a better
conceptual and structural understanding of general quantum field 
theory \cite{Bert97,Morchio88}.

However, there still exist in the literature 
contradictory, and often misleading, conclusions drawn about to the basic 
properties of some of these models. Generally, references made to 
spaces of states 
associated to Wightman functions of redundant field algebras are at the 
origin of the problem. The two-dimensional massless scalar field is the 
simplest example where controversial statements have arisen in the 
literature (see Ref.\cite{Morchio90} for a list of these statements). 

As pointed out in Refs.\cite{Capri81,Morchio88}, the use of 
bosonization techniques to study two-dimensional quantum field
theories raises some non-trivial and delicate questions  related to
the use of a redundante Bose field algebra. This field
algebra contains more degrees of freedom than those needed for the
description of the model and some care must be taken in order to
construct the Hilbert space associated with the Wightmam functions
that define the model. The lack of sufficient appreciation of the
general mathematical structures involved may give rise to misleading
conclusions about the basic structural aspects and physical
properties of the models.

In Ref. \cite{Morchio88} the vector Schwinger model (VSM) was revisited 
through a rigorous
mathematical approach that takes into account the Hilbert topology
which identifies the corresponding Hilbert space structure associated to the
Wightman functions. Some 
features exhibited in the
standard tratments of the VSM were critized and interpreted as consequences 
of considering the representation (Wightman functions) of a redundant
field algebra.

Using an approach based on the treatment given in 
Ref.\cite{Capri81,Morchio88} for the VSM, in Ref.\cite{Carvalhaes97} the  
basic properties of the anomalous chiral Schwinger model (CSM) were 
analyzed on the basis of the principles of 
the general theory of quantized fields. There it was shown that, by 
exercising a careful control on the construction
of the Hilbert space associated to the Wightman functions that
define the 
model, certain structural properties derived in the preceding literature
do not have a mathematical support. In Ref.\cite{Carvalhaes97} we showed 
that the claimed 
need for a $\Theta$-vacuum parametrization in the anomalous chiral 
model \cite{Carena91}, and the suggested 
equivalence of the VSM  and CSM defined for the regularization 
dependent parameter $a=2$ \cite{Carena91,Shizuya88}, cannot be
established in terms of the intrinsic field algebra. These alleged
properties are 
consequences of the use of a redundant field algebra, resulting from
an improper factorization of the closure of the space of states, and 
cannot be regarded as structural features of the theory. In order to obtain a clear
identification of the 
physical state content of the model, in this approach close attention is paid 
to maintaining a complete control 
on the Hilbert space structure needed for the representation of the intrinsic 
field algebra $\mbox{\boldmath $\Im$}$, generated by the set of fundamental local field 
operators $ \{\bar \psi, \psi, {\cal A}_\mu \} $ whose Wightman
functions define the theory. In Ref. \cite{ABRS97} we extended this
analysis to the two-dimensional chiral quantum cromodynamics (CQCD$_2$).

However, the chiral anomalous model is in some sense rather
restricted, since only one fermion field component couples with the 
gauge field. In order to perform a systematic and complete analysis of the 
general basic structural 
aspects of anomalous
two-dimensional gauge theories, in the present work we make an application
of the approach used in Refs. \cite{Morchio88,Carvalhaes97} to the 
so-called generalized 
chiral Schwinger model \footnote{\it Sometimes referred to as
generalized Schwinger model (GSM) \cite{Alvaro94,Naon90}. However, for
reasons which
will become clear along the paper,  and also to stress that we point
out restrictions to the interpretation of the generalized theory as an interpolating 
quantum field theory between pure vector and chiral Schwinger 
models \cite{Basseto94},  we shall refer to it as GCSM.} (GCSM), a 
two-dimensional field theory where left and right fermions couple to the 
gauge field with different 
strengths \cite{Shizuya88,Chanowitz86,Basseto94,Boyanovsky88,Halliday86}. The
generalized model allows an investigation on the existence of the
limit to the pure vector model, usually done in the 
literature \cite{Basseto94,Alvaro94,Naon90,Yan96}.

 Despite
the large amount of papers on the subject, the general mathematical 
structures involved in the two-dimensional anomalous gauge theories 
have not been fully appreciated with the required mathematical
care, and such an analysis seems to be lacking in the existing literature. It is worth and 
also seems to be very instructive to review the generalized chiral
model through a foundational investigation in order to examine some 
delicate mathematical questions that arise from
the use of a redundant field algebra in deriving basic properties of
the model. To this end, in the present paper we employ 
the same systematic treatment 
of Refs. \cite{Morchio88,Carvalhaes97} to perform a careful
discussion of the identification of the physical space content of the
generalized model and study the different conclusions one 
may reach in respect to the physical properties of the model by restricting or 
not to the representation of the intrinsic field algebra. In this approach the states that
are admissible in the Hilbert space $ {\cal H} $ are selected
according to some appropriate criteria, and the physical state
content of the theory is based on the Wightman functions
representing the intrinsic field algebra which describes the degrees
of freedom of the model. For the benefit of the reader we try to 
keep the presentation close to the
one of Ref. \cite{Carvalhaes97}, such that the effect of the right 
and left couplings can be clearly seen.

The paper is organized as follows. In section II, we review the
the operator solution, via bosonization,  of the gauge noninvariant 
formulation ($GNI$) of the anomalous GCSM. In section III we discuss
the cluster decomposition property. In section IV we analyze the
intrinsic field algebra, and the general 
mathematical structures involved in the construction of the Hilbert space 
are discussed. The topological
charge content of the Hilbert space is displayed in order to examine the
effect of the use of a 
redundant Bose field
algebra in deriving basic properties of the model. We show that
different results may arise, as regards the physical properties of
the anomalous model, in restricting or not the
Hilbert space as representation of the intrinsic local field algebra.
In section V the
question related with the VSM as a limit of the GCSM is discussed. We 
argue that in order for
this kind of limit to be fully accomplished for the field algebra (and
for the corresponding Wightman functions) it must be accompanied 
by a singular operator gauge transformation. The vector limit 
changes the algebraic constraint structure of the theory and 
can only be consistently defined for a subalgebra
of the generalized model, so that, rigorously, we cannot 
interpret the GCSM as an interpolating quantum field theory between the 
pure vector and the chiral 
Schwinger models. In Section VI we  briefly extend
the analysis to the discussion of the isomorphic gauge invariant 
formulation ($GI$) of the generalized model. This will also serve to 
streamline the presentation of ref.\cite{Boyanovsky88}. In section VII we present our 
final remarks and conclusions.

\section{Operator Solution: Bosonization}

The generalized chiral Schwinger model (GCSM) is defined from the 
classical Lagrangian density\,\footnote{ {\it The 
conventions used are}: $ \psi = (\psi _{_{\!\!r}},\psi_{_{\!\!\ell}})^{\,T} , \epsilon ^{0 1} = g^{0 0} = - g^{1 1} = 1 $,\\
\centerline{$ \tilde \partial ^{\mu } \equiv  \epsilon ^{\mu \nu} 
\partial _{\nu} , x^\pm = x^0 \pm x^1  , \gamma ^{\,\mu }\gamma ^{\,5}\,=\,
\epsilon ^{\,\mu \,\nu} \,\gamma _{\nu } $,}

\vspace{0.5cm}
\centerline{$ \gamma ^{\,0}\,=\,\pmatrix {0&1 \cr 1&0} \,\,,\,\, 
\gamma ^{\,1}\,=\,\pmatrix {0&-1 \cr 1&0} \,\,,\,\, \gamma ^{\,5}\,=\,
\pmatrix {1&0 \cr 0&{-1}}\,\,.$}} \cite{Boyanovsky88}
\\
$${\cal L} = -\frac{1}{4} ({\cal F}_{\mu\nu})^2 +
i\overline{\psi}\dslash\psi+\frac{1}{2}e_{_r}\overline{\psi}\gamma^\mu(1+\gamma^5)\psi
{\cal A}_\mu + \frac{1}{2}e_{_\ell}\overline{\psi}\gamma^\mu(1-\gamma^5)\psi
{\cal A}_\mu, \eqno (2.1)$$
\\ 
where ${\cal F}_{\mu\nu}$ denotes the usual field-strength tensor. The 
Lagrangian density of the CSM is obtained by taking $e_{_r} = 0$ (or 
$e_{_\ell} = 0$). Although the Lagrangian density of the VSM can 
be ``formally'' obtained from (2.1) in a naive way by 
taking $e_{_\ell} = e_{_r}$, it will be seen that this 
limit can only be consistently defined at the level of a quantized 
field theory if taken for a field
subalgebra of the generalized model.

The classical Lagrangian density (2.1) exhibits invariance under the 
chiral local gauge transformations
\\
$$\psi(x)\rightarrow e^{\frac{i}{2}\left[e_{_r}(1+\gamma^5)+e_{_\ell}
(1-\gamma^5 )\right]\Lambda(x)}\psi(x), \eqno(2.2a)$$

$${\cal A}_\mu(x)\rightarrow {\cal A}_\mu(x) + \partial_\mu\Lambda(x). 
\eqno (2.2b)$$
\\ 

At the quantum level the chiral anomaly spoils the gauge invariance
\cite{Jackiw85,Boyanovsky88} and the effective bosonized theory is given in terms of the Lagrangian 
density \cite{Chanowitz86,Boyanovsky88}
\\
$${\cal L} = -\frac{1}{4}({\cal F}_{\mu\nu})^2+\frac{1}{2}(\partial_\mu\phi)^2 
+ \frac{1}{\sqrt{\pi}}(g_{_-}\partial^\mu\phi+g_{_+} \tilde \partial^\mu\phi)
{\cal A}\mu + \frac{ag_{_+}^2}{2\pi}{\cal A}_\mu{\cal A}^\mu, \eqno(2.3)$$
\\ 
where $ g_{_\pm} = \frac{1}{2}(e_{_\ell} \pm e_{_r}) $ and $a$ plays
the role of the
Jackiw-Rajaraman (JR) parameter that characterizes the ambiguity in
the quantization of the model \footnote{\it The connection with the
regularization used in Ref. \cite{Boyanovsky88} is obtained by setting $ M^2
= a g^2_+$.} (regularization ambiguity of the fermionic 
determinant) \cite{Jackiw85}. The formal equations of motion are
\\
$$\Box\phi + \frac{1}{\sqrt{\pi}}(g_{_-}\partial^\mu{\cal A}_\mu +
g_{_+}\tilde{\partial}^\mu{\cal A}_\mu)=0,\eqno(2.4a)$$

$$\partial_\mu{\cal F}^{\mu\nu}= {\cal J}^\nu = -
\frac{1}{\sqrt{\pi}}(g_{_-}\partial^\nu\phi+g_{_+}\tilde{\partial}^\nu\phi)-
\frac{ag_{_+}^2}{\pi}{\cal A}^\nu.\eqno(2.4b)$$
\\
Introducing into (2.3) a local gauge decomposition for the 
field $ {\cal A}_\mu $,
\\
$${\cal A}^\mu = \partial^\mu\lambda + \tilde\partial^\mu\chi,\eqno(2.5)$$
\\ 
the resulting bosonized effective Lagrangian density can be
treated as a higher-derivative field theory \cite{Halpern}. Following
the procedure used in Refs. \cite{Halpern,Boyanovsky88,Carvalhaes97} we
make use of an {\it enlarged bosonization scheme} and introduce the 
field transformations
\\
$$\phi = \phi^\prime + \frac{1}{\sqrt{\pi}}(g_{_+}\chi-g_{_-}\lambda),
\eqno(2.6a)$$

$$\lambda = \lambda^\prime - \frac{g_{_-}g_{_+}}{(ag_{_+}^2 - 
g_{_-}^2)}\chi,\eqno(2.6b)$$

$$\chi_{_1} = \frac{1}{m}\Box\chi,\eqno(2.6c)$$

$$\chi_{_2} = \frac{1}{m}(\Box+m^2)\chi,\eqno(2.6d)$$
\\ 
where  
\\
$$m^2 = \left( \frac{ag_{_+}^2}{\pi} \right) \frac{ag_{_+}^2-g_{_-}^2+g_{_+}^2}
{ag_{_+}^2-g_{_-}^2}.\eqno(2.6e)$$
\\ 
In this way, the effective Lagrangian density (2.3) is reduced to a local one
\\
$${\cal L} =
\frac{1}{2}(\partial_\mu\phi^\prime)^2+\frac{1}{2}(\partial_\mu\chi_{_1})^2-
\frac{1}{2}m^2\chi_{_1}^2-\frac{1}{2}(\partial_\mu\chi_{_2})^2+
\frac{1}{2}\frac{(ag_{_+}^2-
g_{_-}^2)}{\pi}(\partial_\mu\lambda^\prime)^2.\eqno(2.7)$$
\\

The dynamically generated mass $m^2$ for the field $\chi_{_1}$ and 
the metric sign for the field $\lambda^\prime$ depend 
on the values of the parametric space
variables $(a,g_+,g_-)$. In order to avoid the presence of 
tachyon excitations, the mass $m^2$ must be strictly 
positive and the JR parameter $a$ restricted to two ranges ($g_- \neq
0$)\cite{Basseto94}:
\\
$$(i) \,\,\,a > \frac{g_{_-}^2}{g_{_+}^2} \eqno(2.8a)$$
\\
$$(ii) \,\,\,0 < a < \frac{g_{_-}^2}{g_{_+}^2} - 1\,\,\,
\mbox{ if }\,\,\,\frac{\vert g_{_-}\vert}{\vert g_{_+}\vert}  >  1 \,\,\,\,
\mbox{ or }\,\,\,\, \frac{g_{_-}^2}{g_{_+}^2}-1 < a < 0\,\,\,\mbox{ if }\,\,\,
\frac{\vert g_{_-}\vert}{\vert g_{_+}\vert} <  1. \eqno(2.8b)$$
\\
The dynamics for the field $\lambda^\prime$ is ensured in both
ranges ($ag_{_+}^2-g_{_-}^2 \ne 0$) and no restriction is entailed 
concerning  the dynamics of the field $\phi^\prime$, which describes the free
canonical fermion 
degrees of freedom . In the first range only the massless free 
field $\chi_{_2}$ must be quantized with negative metric, whereas in the 
second one, besides $\chi_{_2}$, $\lambda^\prime$ is also quantized with negative 
metric. As stressed in Refs. \cite{Boyanovsky88,Carvalhaes97}, in the  VSM the gauge 
invariance ensures that the field
$\lambda ^\prime$ is a pure gauge excitation and does not appear in
the corresponding effective bosonized theory. However, in the anomalous chiral
model the additional degree of freedom $\lambda ^\prime$ is a
dynamical field and, as we shall see also in the generalized model,
its presence ensures the existence of fermions in the asymptotic
states, implying that the screening and confinement aspects exhibited
by the anomalous generalized chiral model differ from those of the VSM. As a
matter of fact, the nontrivial anomalous nature of the 
field $ \lambda ^\prime $ ensures the dynamics for the Wess-Zumino
(WZ) field \cite{Carvalhaes97,Belvedere95}.

For $g_- = 0$,  we must have $ a > - 1 $ in order to avoid the presence
of tachyon excitations. In the range $ - 1 < a < 0 $ the field
$\lambda^\prime$ is quantized with negative metric.

The local gauge solutions of the field equations are constructed from the
set of Bose fields $\{\chi_{_1},\chi_{_2},\phi^\prime,
\lambda^\prime\}$, and up to a Klein transformation are given by \cite{Boyanovsky88}
\\
$$\psi_\alpha(x) = :\psi_\alpha^{^o}(x) \exp\left\{-i(g_{_-} -
\gamma^5_{\alpha\alpha} g_{_+}) 
\left[\frac{1}{m}\,(1 - b \gamma^5_{\alpha\alpha}\,) 
\Bigl ( \chi_{_2}(x)-\chi_{_1}(x) \Bigr ) + \gamma^5_{\alpha\alpha} 
\lambda^\prime(x) \right]\right\}:\,,\eqno(2.9a)$$

$${\cal A}^\mu(x) =
\frac{1}{m}\left\{\tilde{\partial}^\mu(\chi_{_2}(x)-\chi_{_1}(x)) -
 b \,\partial^\mu(\chi_{_2}(x)-
\chi_{_1}(x))\right\}+\partial^\mu\lambda^\prime(x),\eqno(2.9b)$$
\\
with $ b = g_{_-}g_{_+} / ( ag_{_+}^2 -g_{_-}^2 ) $. Introducing the two 
independent and well-defined right and left mover fields \cite{Morchio90} 
that split the local massless free scalar 
field $ \phi^\prime (x) = \phi^\prime_{_\ell}(x^+) +
\phi^\prime_{_r}(x^-) $, with a Fock space decomposition $ {\cal
H}_{\phi^\prime} = {\cal H}_{\ell} \otimes {\cal H}_{r} $, the 
components $ \psi_{_{\ell,r}}^{^o} (x) $ of the free and 
massless canonical fermion field 
operator are given by  
\\
$$\psi_{_{\ell,r}}^{^o}(x) = \left( \frac{\mu_{_0}}{2\pi} \right)^{1/2} 
:\,\exp \Big \{ \, \mp 2 i\sqrt{\pi} \phi^\prime _{_{\ell,r}}(x) \Big
\}: \,,\eqno(2.9d)$$
\\
in which we have suppressed the Klein transformations, since they are
not needed for our present purposes. The Wick exponential $ :\exp i
\Phi (x): $ has to be understood as a formal series of Wick-ordered
powers of the field $ \Phi (x) $ at the exponent \cite{Pierotti88}.

The set of field operators corresponding to the operator solution of the 
CSM is obtained
from (2.9) by taking $g_{_-} = g_{_+}$ (or $g_{_-} = -g_{_+}$).

The vector current coupled to the gauge field is obtained from (2.4b)
and (2.9b), and is given by
\\
$${\cal J}^\mu = m\tilde{\partial}^\mu\chi_{_1}+L^\mu,\eqno(2.10)$$
\\ 
where $ L ^\mu $ is a longitudinal current 
\\
$$L^\mu = -
\frac{1}{\sqrt{\pi}}\left\{(g_{_-}\partial^\mu+g_{_+}\tilde{\partial}^\mu)
\phi^\prime
+ \frac{1}{\sqrt{\pi}}\left[(ag_{_+}^2-g_{_-}^2)\partial^\mu\lambda^\prime-
g_{_-}g_{_+}\tilde{\partial}^\mu\lambda^\prime\right] + 
m\sqrt{\pi}\tilde{\partial}^\mu\chi_{_2}\right\} \eqno(2.11) $$
\\ 
satisfying
\\
$$ \bigl [\,L^{\,\mu }(x)\,,\,L^{\,\nu }(y)\,\bigr ]\,=\,0\,\,
\,,\forall \,x,y\,\,, \eqno (2.12) $$
\\ 
such that it generates, from the vacuum $ \mbox{\boldmath{$\Psi_o$}} $, zero norm states
\\
$$ \langle \,L_{\mu }(x)\,L_{\nu }(y)\,\rangle_o \equiv \langle \,
L_{\mu }(x)\,\mbox{\boldmath{$\Psi_o$}}\,,\,L_{\nu }(y)\,\mbox{\boldmath{$\Psi_o$}}
\,\rangle\,=\,0\,\,. \eqno (2.13) $$
\\

For further convenience, let us {\it enlarge the algebra of the Bose
fields} by introducing the set of dual 
fields $ \tilde \Phi = \{ {\tilde \phi} ^\prime,
 \tilde {\lambda ^\prime}, \tilde \chi _{_2} \} $ such
that $ \partial _\mu {\tilde \Phi } = - \epsilon _{\mu \nu} \partial
^\nu \Phi  $. In this way, the longitudinal current $ L_\mu $ can be
written in terms of the scalar potential $ L $ and 
pseudo-scalar potential  $ \tilde L $ as
\\
$$L^\mu=-\frac{1}{\sqrt{\pi}}\,\partial^\mu L = 
\frac{1}{\sqrt{\pi}}\,\tilde{\partial}^\mu\tilde{L}\,,\eqno(2.14)$$
\\ 
where
\\
$$L_{_r} = \frac{1}{2}(L+\tilde{L})= (g_{_-}-g_{_+})\,\phi^\prime_{_r} +
\frac{1}{\sqrt \pi}(ag_{_+}^2-g_{_-}^2+g_{_-}g_{_+})\,\lambda^\prime_{_r}
- m\sqrt{\pi}\,\chi_{_{2r}},\eqno(2.15a)$$
\\
$$L_\ell = \frac{1}{2}(L-\tilde{L})= (g_- + g_+)\phi^\prime_{_\ell} +
\frac{1}{\sqrt \pi} (ag_{_+}^2-g_{_-}^2-g_{_-}g_{_+})
 \lambda^\prime_{_\ell}+ m\sqrt{\pi} \chi_{_{2\ell}}.\eqno(2.15b)$$
\\

As is well know, due to the presence of the longitudinal current $
L_\mu $,  Gauss's law holds in a weak form and is satisfied on
the physical subspace ${\cal H}_{phys}$ defined by the constraint $
L_{\,\mu} \approx 0 $.

\section{Cluster Decomposition Property}

The operator solution is given in terms of free fields and the 
general Wightman functions of the model can be easily computed by
normal-ordering any product of the fundamental fields $\{\bar
\psi,\psi,{\cal A}_\mu\}$. In order
to display the occurence of asymptotic factorization we consider for simplicity
the fermionic two-point functions, which are given in terms of the
corresponding free functions by
\\
$$\langle \psi_{_r}^\ast(x)\,\psi_{_r}(0) \rangle = \langle \psi_{_r}^{o\,\,
\ast}(x^+)\,\psi_{_r}^o(0) \rangle\,\times $$
\\
$$ \exp \Bigg \{
\frac{ (g_{_-}^2-g_{_+}^2)^2}
{m^2(ag_{_+}^2-g_{_-}^2)} \Delta^{^{(+)}}(x,0) \Bigg \}\,\exp \Bigg \{
\frac{(g_{_-}-g_{_+})^2}{m^2}
\Bigl [1-\frac{g_{_-}g_{_+}}{ag_{_+}^2-g_{_-}^2} \Bigr ]^2\,\Delta^{^{(+)}}
(x,m) \Bigg \}, \eqno(3.1a)$$
\\
$$\langle \psi_{_\ell}^\ast(x)\,\psi_{_\ell}(0) \rangle = \langle 
\psi_{_\ell}^{o\,\,\ast}(x^-)\,\psi_{_\ell}^o(0) \rangle \,\times $$
\\
$$ \exp \Bigg \{ \frac{ 
(g_{_-}^2-g_{_+}^2)^2}{m^2(ag_{_+}^2-g_{_-}^2)}
\Delta^{^{(+)}}(x,0)\Bigg \}\,\exp \Bigg \{ \frac{(g_{_-}+g_{_+})^2}{m^2} 
\Bigl [1+\frac{g_{_-}g_{_+}}{ag_{_+}^2-g_{_-}^2}
\Bigr ]^2\,\Delta^{^{(+)}}(x,m) \Bigg \}. \eqno(3.1b)$$
\\ 
For long distances, the contributions coming from the 
free-fermion two-point functions are weighted by the
contributions of the two-point functions of the  
fields $\chi_{_2}$ and $\lambda^\prime$. For $g_{_-} = g_{_+}$ 
($g_{_-} = -g_{_+}$) the effects mediated by the 
massless fields $ \lambda^\prime $ and $ \chi_{_2} $ disappear and we 
recover the correlation functions of the CSM \cite{Carvalhaes97}.

In the first range the correlation functions (2.16) exhibit asymptotic 
factorization expressing statistical independence for long distances.
As we shall see, in the first range the cluster decomposition is not violated 
by the 
Wightman functions that define the model, implying that there is no need for
a $\Theta$-vacuum parametrization. 

The possibility of considering the second range (2.8b) for the parameter $a$ 
was put forward by Basseto et al. \cite{Basseto94}. At first sight, it 
appears that, with this input, the physical content of the GCSM is more 
abundant than it was supposed before \cite{Yan96}. However, we do not
think that this is the case. Instead, we shall argue below that the theory 
loses most of  its physical meaning in this domain, in spite of the absence of 
tachyons. From our point of view, the first range (2.8a) remains the 
region where the theory is consistently defined.

In the second range, the fields $ \lambda^\prime $ and $ \chi_{_2} $
are quantized with negative metric implying the presence of long
range correlations. These contributions  to the
two-point functions, coming from the first exponential factor in
Eq.(2.16),  increase with the distance as $ x^{2 \vert \gamma \vert} $, with
$ \vert \gamma \vert =  
(g_-^2 - g_+^2 )^2 / ( 4 \pi m^2 \vert ag_+^2 - g_-^2 \vert ) $. However, since
in the second range unitarity is spoiled \cite{Basseto94}, we just discard any
attempt to interpret this ``phenomena'' as a signal of confinement.

The solution given in Ref. \cite{Basseto94} is constructed \footnote{\it The 
mapping from the parameters space $(a,g_+,g_-)$ to $(a,r)$, considered 
in Ref. \cite{Basseto94}, is obtained by 
taking $ e_\ell \equiv  e ( 1 + r ) $ , $ e_r \equiv e ( 1 - r ) $,
where the parameter $r \in [ - 1, 1 ] $.} only in 
terms of two Bose fields, the massive 
field $\sigma \equiv - (m \sqrt\pi / g_+ )\,\chi_1 $, and a massless 
field $ h $, which includes the free-fermion degrees of freedom. In the 
second range the field $h$ is quantized with negative metric, which 
implies that unitarity is threatened. In 
order to remove this ``ghost'' and ``restore'' unitarity, in Ref. \cite{Basseto94} the 
subsidiary condition $ h^{(+)}(x) \vert \Psi_{phys} \rangle = 0 $ is 
imposed, which 
consists in extracting the field $h$ from the 
observable algebra in the second range. This 
condition, however, is too restrictive and implies, for example, that 
the free fermion degree of freedom is extracted  from 
the theory, as
well as that the 
fermion operator and the gauge field fail to be physical, contrary to
what is expected in an anomalous gauge theory \cite{Carvalhaes97,ABRS97}. 

Since in the second range the correlation functions (3.1) increase 
with the distance, this behavior  was 
interpreted in Ref. \cite{Basseto94} as confinement. Nevertheless, in the 
approach of Ref.\cite{Basseto94} the fermion field is unphysical in the 
second range (it does not commute with $h$). This interpretation 
looks, therefore, meaningless. Because of the 
loss of physical field status by the fermion field operator, the 
statement made in \cite{Basseto94} that the electric charge is 
totally screened due 
to the neutrality of $\psi$ lacks physical meaning. Furthermore, upon 
excluding the field $h$ from the field algebra, the authors 
of Ref.\cite{Basseto94} have
restricted themselves to a subspace of the Hilbert space of the theory, namely 
to the Fock space of the scalar massive field $\sigma$. Therefore, the 
physical properties derived in this framework do not correspond to those 
of the original anomalous gauge theory.

\section{Intrinsic Local Field Algebra}

Following the treatment of Refs. \cite{Morchio88,Carvalhaes97}, in this 
section we shall undertake a careful 
analysis of the Hilbert space structure and general properties of the 
anomalous GCSM. We restrict the discussion only to the first range of the 
parameter $a$, for which  unitarity is not violated.  The
procedure which we shall adopt to display the basic structural
properties of the model and obtain a consistent prescription to
identify its physical state content,  following the general strategy
introduced in Ref. \cite{Capri81,Morchio88}, is to embed the mathematical
structures of the bosonization scheme into the context of the general
principles of Wightman field theory. In this way the states that are
admissible in the Hilbert space $ {\cal H} $ are selected according
to some appropriate criteria and the physical interpretation of the
theory is based on the Wightman functions representing the intrinsic
local field algebra which describes the degrees of freedom of the model.

Within the treatment of the model in a local gauge formulation we have lack of
positivity of the Wightman functions leading to an indefinite metric
field theory. The strategy is then to consider the solution of the model
satisfying all the Wightman axioms except positivity, i. e., the
modified Wightman axioms including the Hilbert space structure condition
\cite{Strocchi74,Morchio80,Morchio87,Morchio88}. As pointed out in
Ref. \cite{Morchio88}, within the approach based on the intrinsic
field algebra $ \mbox{\boldmath $\Im$} $ generated by the set of local field 
operators $ \{\bar \psi,\psi,{\cal A}_\mu \} $, the prescription to
identify correctly the physical state content of the model is to
consider a Hilbert space of states $ {\cal H} $ obtained by
completion of the local states according to a suitable
minimal Hilbert topology. This enables the identification of a Hilbert
space structure associated with the Wightman functions of $ \mbox{\boldmath $\Im$} $ and
define the theory. The last step is to 
impose the subsidiary (Gupta-Bleuler-like) condition to get the 
physical Hilbert subspace \footnote{\it For a more detailed discussion see
Refs.\cite{Morchio90,Capri81,Morchio88,Carvalhaes97}.}.

In this approach, the 
basic object is the set of field 
operators $ \{ \bar \psi,\psi , {\cal A}_\mu \} $, which are the elements of the
local field algebra intrinsic to the model, and generates, through 
linear combinations, polynomials of these smeared fields,  Wick products, point-splitting 
regularizations of polynomials, etc., an intrinsic local field algebra $\mbox{\boldmath $\Im$}$. These 
field operators constitute the intrinsic mathematical description of
the model and serve as a kind of building material in terms of which the 
model is formulated 
and whose Wightman functions define the model. Every operator of the
theory is a
function of the intrinsic set of field operators $ \{\bar \psi, \psi,
{\cal A}_{\mu}\} $ which defines a polynomial 
algebra $ \mbox{\boldmath $\Im$} = {\bf \mbox{\LARGE{\boldmath $\wp$}}} (\{\bar \psi,\psi,{\cal A}_\mu\}) $. The local
field algebra $ \mbox{\boldmath $\Im$} $ identifies a 
vector space ${\cal D}_{_0}=\mbox{\boldmath $\Im$} \mbox{\boldmath{$\Psi_o$}}$ of local states 
where the inner product is defined by Wightman functions of $\mbox{\boldmath $\Im$}$. 

The effective (bosonized) theory is formulated in terms of the 
set of Bose field 
operators $\{\phi^\prime , \chi_{_1} , \chi_{_2} ,  \lambda^\prime\}$
and using a Fock vacuum for them. This set of 
fields defines a largely redundant local algebra $\mbox{\boldmath $\Im$}_{_B}$ generated 
through Wick exponentials and derivatives of these Bose fields. The 
algebra $\mbox{\boldmath $\Im$}$ is a proper subalgebra of the local 
algebra $\mbox{\boldmath $\Im$}_{_B}$. It follows that, due to the nonpositivity coming
from the two-point function of the massless scalar field $ \chi_{_2}
$,  the field algebra $\mbox{\boldmath $\Im$}_{_B}$ defines a vector 
space ${\cal D}_{_0}^\prime = \mbox{\boldmath $\Im$}_{_B} \mbox{\boldmath{$\Psi_o$}}$ with indefinite 
metric and that contains elements which are not intrinsic 
to the model. The strategy to associate a Hilbert space of 
states to the Wightman 
functions of $\mbox{\boldmath $\Im$}_{_B}$ consists in considering \cite{Morchio88} the Hilbert 
completion ${\cal K}_{_B}$ of ${\cal D}_{_0}^\prime$ 
according to a suitable minimal 
Hilbert topology $ \tau $, namely the Krein topology associated to the 
Wightman functions 
of $\mbox{\boldmath $\Im$}_{_B}$: ${\cal K}_{_B} = 
{\overline{{\mbox{\boldmath $\Im$}}_{_B}\,\mbox{\boldmath{$\Psi_o$}}}} ^{\,\tau}$. In this 
construction it results that the  closure 
of ${\cal D}_{_0}$  in a Hilbert topology $\tau$ 
defines \cite{Morchio88,Morchio90} a 
Hilbert space $ {\cal H} = {\overline{\mbox{\boldmath $\Im$}\, \mbox{\boldmath{$\Psi_o$}}}}^{\, \tau} $, in which the 
algebra $\mbox{\boldmath $\Im$}$ is represented. The Hilbert space $\cal H$, which
provides a representation of the field algebra $ \mbox{\boldmath $\Im$} $,  is then a proper 
subspace of the Krein space ${\cal K}_{_B}$, which provides a Fock-Krein
representation of the algebra $ \mbox{\boldmath $\Im$}_{_B} $.

Setting $ {\cal H} = \overline{\mbox{\boldmath $\Im$} \mbox{\boldmath{$\Psi_o$}}}^{\,\tau} $, and since 
the 
field $ \chi_{_1} = 1 / 2 m \varepsilon^{\mu \nu} {\cal F}_{\mu \nu} \in \mbox{\boldmath $\Im$} $ 
is massive, in analogy with the VSM \cite{Morchio88} the 
Hilbert space $\cal H$ can be decomposed as a tensor 
product 
\\
$$ {\cal H} = {\cal H}_{\chi_{_1}} \otimes {\cal H}_{\chi_{_2},
\lambda^\prime, \psi_o}\,,\eqno (4.1) $$
\\
where $ {\cal H}_{\chi_{_1}} $ is the Fock space 
of the massive field $\chi_{_1}$, and the closure of the space is
\\
$$ {\cal H}^\prime_{\chi_{_2}, \lambda^\prime, \psi_o} = 
{\overline{\mbox{\boldmath $\Im$}^\prime_{\chi_{_2},\lambda^\prime,\psi_o} \mbox{\boldmath{$\Psi_o$}}}}^{\,\tau} \,,
\eqno (4.2)$$
\\
where $ \mbox{\boldmath $\Im$}^\prime_{\chi_{_2},\lambda^\prime,\psi_o} $ is the
local field subalgebra not containing $\chi_{_1}$ and 
generated by $ L^\mu $ and the Wick exponentials
of the massless fields appearing in Eqs.(2.9a), and which commute 
with $L_\mu$. In analogy with the VSM \cite{Morchio88}, the massive
field $\chi_{_1} $, as well as the Wick exponential 
operators \cite{Pierotti88}
\\
$$ G_\alpha (x) \doteq\, : e^{\,\frac{i}{m}\,
(g_-\,+\,\gamma_{\alpha \alpha}^5\, g_+)
(1\,+\,b\,\gamma_{\alpha\alpha}^5) \chi_{_1}(x) } :\,,\eqno (4.3)$$
\\ 
which appear in the expression for the fermion field 
operator (2.9a), are elements of the intrinsic 
field algebra $\mbox{\boldmath $\Im$}$. This enables the definition of the field operator
\\
$$ \psi^\prime (x) \doteq\, : G^\ast(x) \psi (x) :\,,\eqno (4.4)$$
\\
as an element of the field subalgebra $ \mbox{\boldmath $\Im$}^\prime \subset \mbox{\boldmath $\Im$}$. As
we shall see, the operator $\psi^\prime$ cannot be reduced and the
closure of the 
space ${\cal H}^\prime_{\chi_{_2}, \lambda^\prime,\psi^o}$ cannot be 
further decomposed. Except, for the chiral case $ g_+ = g_- $ ($ g_+
= - g_- $), in which the right-(left)-fermion component is free, the
Hilbert space can be decomposed as 
\\
$$ {\cal H} = {\cal H}_{\chi_{_1}}
\otimes {\cal H}^\prime_{\chi_{_2}, \lambda^\prime, \psi^o_\ell}
\otimes {\cal H}_{\psi^o_r}\,.\eqno (4.5)$$
\\
The improper decomposition of the 
closure of the
space ${\cal H}^\prime$ implies the use of a redundant field algebra, and 
leads to physical consequences which do
not correspond to intrinsic basic properties of the model.

With the introduction of the set of dual field 
operators $ \tilde \Phi = \{ \tilde{\chi}_{_2} , 
\tilde{\lambda}^\prime \} $, the algebra $\mbox{\boldmath $\Im$}_{_B}$ is      
enlarged, generating an external 
algebra $\mbox{\boldmath $\Im$}^{\,ext}_{_B}$, which contains redundant elements not
intrinsic to the model. Consequently, $\mbox{\boldmath $\Im$}_{_B}$ is a proper subalgebra 
of $\mbox{\boldmath $\Im$}^{\,ext}_{_B}$. The algebra $ \mbox{\boldmath $\Im$}^{\,ext}_{_B} $ defines the 
Krein space ${\cal K}^{\,ext}_{_B} = \overline{\mbox{\boldmath $\Im$}^{\,ext}_{_B}\mbox{\boldmath{$\Psi_o$}}}$, which 
properly contains ${\cal K}_{_B}$: $ {\cal K}_{_B}^{ext} \supset {\cal
K}_{_B} \supset {\cal H} $.

The appearance of the longitudinal current $L^\mu$ in Eq.(2.4b) implies the 
weakening of the local Gauss's law, i.e.,
\\
$$\Bigl \langle \Phi , \Bigl ( J^\nu(x) - \partial_\mu {\cal
F}^{\mu\nu}(x) \Bigr ) \Psi \Bigr \rangle = 
\Bigl \langle \Phi , L^\nu(x) \Psi \Bigr \rangle = 0\,. \eqno (4.6) $$
\\ 
The selection of physical states by means of the subsidiary
condition $ L_\mu \approx 0 $, must be correctly done by specifying the 
Hilbert space on which such condition is imposed. Let
\\
$$ \hat {\cal H} \equiv \Bigl \{ \vert \Phi \rangle\, \in\, {\cal H}\,
 \Big \vert\, (L^{(-)}_\mu  ) \,\vert \Phi \rangle\, =\, 0 \Bigr \}\,,
\eqno (4.7) $$
\\
be the subspace of solutions obeying this condition. To get a physically
acceptable interpretation of the model, one has to specify the space
of physical states on which Maxwell's equations hold as operator
equations, i. e. ,
\\
$$ {\cal H}_{phys} = {\overline{\hat {\cal H} \Big / \hat {\cal H}_o}}
^{\hat \tau}\,, \eqno (4.8) $$
\\
where $ \hat \tau $ is the induced topology on the quocient and $
\hat {\cal H}_o $ is the null subspace of $ \hat {\cal H} $.

The 
algebra of physical observables $\mbox{\boldmath $\Im$}_{phys}$ is generated from the 
set of operators $ \{ {\cal O} \} \in \mbox{\boldmath $\Im$}$ under whose 
applications $\hat {\cal H}$ is stable, 
\\
$$ \vert \Psi \rangle \in \hat {\cal H} \Rightarrow {\cal O} 
\vert \Psi \rangle \in \hat {\cal H}\, .\eqno(4.9)$$
\\ 

In contrast to what happens in a genuine gauge theory (such as the
VSM), {\it in the anomalous model all 
operators belonging to the intrinsic local field algebra $\mbox{\boldmath $\Im$}$ satisfy 
the subsidiary condition, and thus represent physical 
observables of the theory, and the physical 
space $ \hat {\cal H}$ is identified with the Hilbert space $ {\cal H}$}. The 
fact that the intrinsic set of field 
operators $ \{\bar \psi,\psi,{\cal A}_\mu \} $ represents physical
observables of the GCSM gives rise to a basic structural 
distinction between the
anomalous theory and the genuinely gauge-invariant one.

However, for special values of the parametric space variables,  delicated
questions may arise concerning the structural properties of the
anomalous models and the determination of their physical content. In 
this case, the bosonization procedure 
introduces a broader class of operators belonging to the field algebras
$\mbox{\boldmath $\Im$}^{^{ext}}_{_B}$ , $\mbox{\boldmath $\Im$}_{_B}$ which commute with the longitudinal
current $L_\mu$. These are redundant operators that do not represent
intrinsic elements of the field algebra $\mbox{\boldmath $\Im$}$ and their appearance requires 
an additional care in the construction of the Hilbert space $ {\cal H} $.

Since the physical content of the anomalous models relies
strongly and directly on the intrinsic field algebra, in order to get 
unequivocal conclusions the general structural 
properties of the model must be analyzed
taking a careful control on the Hilbert space associated with the
Wightman functions that provide a representation of the field algebra
generated from the set of the fundamental field operators of the
model. In order to implement the strategy of filtrating from the bosonized theory the
elements intrinsic to the field algebra $\mbox{\boldmath $\Im$}$, we must analyze the 
charge content of the Hilbert space $ {\cal H} $. From the fact that some
{\it topological} charges get trivialized in the restriction from $
{\cal K} ^{^{ext}}_{_B} $ or $ {\cal K}_{_B} $ to $ {\cal H} $, one
infers that the
closure of local states associated with the intrinsic field algebra
does not allow the introduction of operators that carry these
trivialized topological charges \cite{Morchio88,Carvalhaes97}. In order 
to show that by relaxing this careful construction of the physical 
Hilbert space of the anomalous model
one can lose the complete control on the corresponding field theory and
some non-trivial and delicate mathematical questions can arise, it appears 
instructive to examine the effect of the use
of a redundant field algebra in deriving basic properties of
the model. As we shall see, different results may arise, in respect
to the physical properties of the generalized anomalous model, in
restricting or not to the representation ${\cal H}$ of the intrinsic
local field algebra $ \mbox{\boldmath $\Im$} $.

To begin with, we shall consider the model defined in the first range
and in the particular case 
with $ ag_{_+}^2 - g_{_-}^2 = g_{_-}g_{_+}$ ( $ g_+ \neq - g_- $ ), in which 
the corresponding  operator algebra exhibits 
delicate special features. This is the generalized version of the 
case $a = 2$ of the CSM which has generated some confusion in the literature
\cite{Carena91,Shizuya88}. Note that, in the first range and
within the branch $ ag_{_+}^2 - g_{_-}^2 = g_{_-}g_{_+} $ 
( $ ag_{_+}^2 - g_{_-}^2 =  - g_{_-}g_{_+} $ ), in the 
limit $ g_+ = g_- $ ( $ g_+ = - g_- $ )  we recover the usual chiral model 
with $ a = 2 $.

In the coupling constant domain $ ag_{_+}^2 - g_{_-}^2 = g_{_-}g_{_+}$, the 
field component $\lambda_\ell^\prime$ decouples 
from the longitudinal current (2.11) and the components of the 
fermion field operator 
are given by ($\varphi_{_{\ell,r}}\equiv \varphi (x^\pm)$) 
\\
$$ \psi_{_r}(x) =  \psi^o_{_r}(x) : e^{ - i ( g_{_-} - g_{_+})
\lambda^\prime(x) }:\,, \eqno(4.10a) $$

$$ \psi_{_\ell}(x) = \left( \frac{\mu_o}{2\pi} \right)^{1/2} 
:e^{2i\sqrt{\pi}\chi_{_1}(x)}: 
\sigma_{_\ell}(x) \Gamma_{_r}(x) \Lambda_{_\ell}(x), \eqno(4.10b) $$
\\ 
where $ \sigma_{_\ell}(x) $, $ \Gamma_{_r}(x) $, $ \Lambda_{_\ell}(x)
$, are operators that commute with $ L_\mu $, and are given by
\\
$$ \sigma_{_\ell}(x) = :e^{-2i\sqrt{\pi} \left (  \phi^\prime_{_\ell}(x) + 
\chi_{_{2l}}(x) \right ) }:,\eqno(4.11a)$$

$$ \Gamma_{_r}(x) = : e^{-2i\sqrt{\pi} \left ( \chi_{_{2r}}(x) - 
\frac{ ( g_{_-} + g_{_+} )}{2\sqrt{\pi}} \lambda^\prime_{_r}(x)
\right ) } :, \eqno(4.11b)$$

$$\Lambda _{_\ell}(x) = : e^{i (g_{_-} + g_{_+}) 
\lambda^\prime_{_\ell}(x)} :. \eqno(4.11c)$$
\\ 
The operator $\sigma_{_\ell}$ is the spurion operator which appears in the
Lowenstein-Swieca solution of the VSM \cite{Lowenstein71,Belvedere79}. The 
operator $ \sigma_{_\ell} $ carries the left free-fermion chirality, commutes with the longitudinal current $ L_\mu $ and
generates constant Wightman functions. Note that despite of being
independent of the negative metric field $ \chi_{_2} $, the field component
$ \psi ^o_{_r} (x) $, given by Eq.(4.10a), commutes with the
longitudinal current. 

For $ g_+ = g_- \equiv g $ we 
obtain from (4.10-11) the operator solution of the 
CSM  with $ a = 2 $ \cite{Carvalhaes97}. In this case we have
\\
$$ \langle \mbox{\boldmath{$\Psi_o$}},\chi_{_{2 r}}(x) \chi_{_{2 r}}(y) \mbox{\boldmath{$\Psi_o$}} \rangle +
\frac{g^2}{4 \pi} \langle \mbox{\boldmath{$\Psi_o$}},\lambda^\prime_{_r}(x) \lambda^\prime_{_r}(y) 
\mbox{\boldmath{$\Psi_o$}} \rangle = 0 \,,\eqno (4.12) $$
\\
and the operator $ \Gamma_{_r} $, given by (4.11b), also becomes a spurious 
operator 
\\
$$ \Gamma_{_r}(x) \equiv \hat\sigma_{_r}(x) = :e^{ 2i\sqrt{\pi} \left [ \chi_{_{2 r}}(x) - 
\frac{g}{ 2 \sqrt \pi} \lambda^\prime_{_r}(x) \right] }:\,.\eqno(4.13) $$
\\
The operator $ \hat \sigma_{_r} $ does not carry the right free-fermion
chirality and we use the ``hat'' notation to distinguish it from the
spurion operator $ \sigma_{_r} $ that appears in the Lowenstein-Swieca 
solution (see eq. (;;;)). The components of the
fermion field operator are given by
\\
$$ \psi_{_r}(x) = \psi ^o_{_r} (x)\,,\eqno (4.14a)  $$
\\
$$ \psi_{_\ell}(x) = \left( \frac{\mu_o}{2\pi} \right)^{1/2} 
:e^{2i\sqrt{\pi}\chi_{_1}(x)}: \sigma_{_\ell}(x) \hat \sigma^\ast_{_r}(x)
:e^{\,i g \lambda^\prime_{_\ell}(x) }:\,, \eqno (4.14b) $$
\\
In this case, the Wightman functions of the 
operator $\psi^\prime$, defined by (4.4) and given by
\\
$$ \psi^\prime_{_\ell} (x) \,\doteq\, \left( \frac{\mu_o}{2\pi} \right)^{1/2}\,
\sigma_{_\ell}(x) \hat \sigma^\ast_{_r}(x)
:e^{\,i g \lambda^\prime_{_\ell}(x) }:\,,\eqno (4.15) $$
\\
which appear in (4.14b), are isomorphic to those of the free and massless 
canonical left fermion field
operator $ \psi_{_\ell}^o $,
\\
$$ \langle\,\prod _{i=1}^n\,{\psi^{\prime}}^*(x_i)\,\prod_{j=1}^n\,\psi^\prime
(y_j)\,\rangle\,\equiv\,
 \langle\,\prod _{i=1}^n\,\psi^{*\,0}_{_{\!\!\ell}}(x_i)\,\prod _{j=1}^n\,
 \psi^{\,0}_{_{\!\!\ell}}(y_j)\,\rangle\,. \eqno (4.16) $$
\\

Within the formulation based on the representation of the intrinsic field 
algebra, the field operator $ \psi^\prime(x) $ cannot be reduced and the 
operators 
$ \sigma_{_\ell} $, $ \hat 
\sigma_{_r} $, $ \sigma_{_\ell}^\ast \hat \sigma_{_r} $ do 
not exist in $ {\cal H} $. In the first range and in the branch $ ag_{_+}^2 - g_{_-}^2 = g_{_-}g_{_+} $ ( $ ag_{_+}^2 -
g_{_-}^2 = - g_{_-}g_{_+} $ ) there is a broader class of operators
belonging to the field algebras $\mbox{\boldmath $\Im$}^{ext}_{_B}$, $ \mbox{\boldmath $\Im$}_{_B} $ which
commute with the longitudinal current $L_\mu$ and it is 
tempting to extract them from the operators (4.11b,4.14b)
and try to define the spurious 
operators $\sigma_{_\ell}$, $\hat \sigma_{_r}$ in $ {\cal H} $, since the dependence on
the massive field $ \chi_{_1} $ can be 
factorized \cite{Morchio88,Carvalhaes97}. However, as we shall see,
the states generated from these operators do not exist as solution of
the subsidiary condition (4.6) in $ {\cal H} $.

For $ g_+ = g_- \equiv g $, and aside from the
Wick exponential of the ``decoupled'' massless field component $
\lambda^\prime_{_\ell} $, the operator (4.14b) corresponds to the composite 
chiral density operator of the VSM \cite{Carvalhaes97,Carena91}
\\
$$ M (x) \doteq \,: \psi_{_\ell}(x)\,e^{\,- i g
\lambda^\prime_\ell(x)}: =\,\left( \frac{\mu_o}{2\pi} \right)^{\frac{1}{2}}         
:e^{2i\sqrt{\pi}\chi_{_1}(x)}: \sigma_{_\ell}(x)
\hat \sigma^\ast_{_r}(x)\,. \eqno (4.17) $$
\\
As shown in \cite{Carvalhaes97}, the Wick exponential
$ :e^{ i g \lambda^\prime_{_\ell}(x) }: $ is not an element of the
intrinsic field algebra $ \mbox{\boldmath $\Im$} $ and thus cannot be extracted 
from (4.14b). The 
operator $ M (x) $, which violates the cluster decomposition property, cannot 
be defined as a solution of the subsidiary condition in $ {\cal H} $. The 
translationally invariant 
condensed states carrying the free-fermion left chirality
($n_{_\ell}$), like $ \vert n_{_\ell} \rangle \equiv \sigma_{_\ell}^n \,
\mbox{\boldmath{$\Psi_o$}} $, belong to the improper Hilbert space decomposition of the 
closure of the space of local states $ {\cal H}^\prime $, such 
as $ {\cal H}^\prime = {\cal H}_{\psi^o_r} \otimes {\cal H}_{\psi^o_\ell}
\otimes {\cal H}_{\chi_{_{2 r}}} \otimes {\cal H}_{\chi_{_{2 \ell}}} \otimes
{\cal H}_{\lambda^\prime_{r}} \otimes {\cal H}_{\lambda^\prime_\ell} $, and 
cannot be regarded as a state in the Hilbert space which defines the 
representation of the intrinsic field 
algebra $ \mbox{\boldmath $\Im$} $ \cite{Morchio88,Carvalhaes97}. As stressed 
in Ref.\cite{Carvalhaes97}, this observation points out restrictions to the
conclusions of Ref. \cite{Carena91} referring to the necessity of a
$\Theta$-vacuum parametrization in the CSM.

In the branch $ a g_+^2 - g_-^2 = g_+ g_- $, and for $ g_+ \neq g_- $, since 
the Wick exponentials $\Lambda_{_\ell}$ and
$\Gamma_{_r}$ commute with the constraints, it is also tempting to extract 
them from the operator (4.10) and try to define the 
operator $ \sigma_{_\ell} $ in $ {\cal H} $. However, as we shall
see,  neither the exponentials $ \Gamma_{_r} $ and $ \Lambda_{_\ell} $ can be defined 
separately in the Hilbert space $ {\cal H} $ nor their product.

Consider the branch $ a g_+^2 - g_-^2 = g_+ g_- $. The massless contribution of the gauge 
field  $ {\cal A}^L_\mu \in \mbox{\boldmath $\Im$}^\prime_{\chi_{_2},\lambda^\prime,\psi^o}
\subset \mbox{\boldmath $\Im$} $, can be written as
\\
$${\cal A}^L_\mu(x) = \frac{1}{e_{_\ell}}\,\partial _\mu  A (x)
\,,\eqno (4.18)$$
\\
with \footnote{ \it The massless scalar field $ A (x) $ can also be
defined as  
\\
$$ A (x) \doteq e_{_\ell} \int_{- \infty}^x {\cal A}^L_\mu (\xi) d \xi^\mu\,, 
$$
by requiring that, in the line integral extending to infinity over the 
space-time derivatives of the massless scalar fields, the boundary conditions 
are choosen to ensure local integrability and to avoid the introduction of 
variables at infinity.}
\\
$$ A (x) = \sqrt \pi \Bigl ( \tilde \chi_{_2} (x) - \chi_{_2}
(x) \Bigr ) + e_{_\ell} \lambda^\prime(x)\,,\eqno (4.19) $$ 
\\
Note that in the anomalous case the massless piece $ {\cal A}^L_\mu
(x) $ commutes with the longitudinal current
\\
$$ \Bigl [ {\cal A}^L_\mu(x) , L_\mu(y) \Bigr ] = 0\,.\eqno (4.20) $$
\\
The operator $ \Gamma_{_r}(x)\Lambda_{_\ell}(x) $ can be written as
the Wick exponencial of the field $ A(x) $
\\
$$\Gamma_{_r}(x)\Lambda_{_\ell}(x) = : e^{\,i A (x) } :\,,\eqno (4.21) $$
\\
and extraction of the operator $ \Gamma_{_r} \Lambda_{_\ell} $ in 
Eq.(4.11b) can be 
performed by defining the formally ``gauge invariant'' field operator
\footnote {\it This operator can also be defined as
$$ \Omega_{_\ell}(x) \doteq \,: \psi_{_\ell}(x)\,e^{\,- i e_\ell
\,\int_{- \infty}^x {\cal A}^L_\mu(\xi) d \xi^\mu } : \,
\equiv \,: e^{\,2\, i \,
\sqrt \pi \,\chi_{_1}(x) }: \sigma_{_\ell} (x)\,. $$}
\\
$$ \Omega_{_\ell}(x) \doteq \,: \psi_{_\ell}(x)\,e^{\,- i A (x)} :
\,. \eqno (4.22) $$
\\
The operator $ \Omega_\ell $ also commutes with the 
longitudinal current
\\
$$ \Bigl [  \Omega_\ell (x) , L_\mu (y) \Bigr ] = 0\,,\eqno (4.23) $$
\\
and due to the presence of the spurious operator $\sigma_{_\ell}$, violates 
the cluster decomposition property.  The ``infrared dressed'' field operator
(4.22) is the (anomalous) analogue of the {\it bleached fields}  of the
Lowenstein-Swieca solution of the VSM
\cite{Lowenstein71,Belvedere79}, whose existence as an element of the
intrinsic field algebra of the gauge invariant vector model has been
critized in Ref. \cite{Morchio88}. 

From this
construction, based on the decomposition of the closure of the space, one 
may  be induced to conclude in favor of the need for a
chiral $\Theta$-vacuum parametrization in the generalized chiral model
defined for this coupling constant branch. However, as we shall see, the 
operator $ \Omega_\ell (x) $ does not belong to the 
representation ${\cal H}$ of the 
intrinsic local field algebra $\mbox{\boldmath $\Im$}$ and thus cannot be defined as a solution
of the subsidiary condition in $ {\cal H} $.

In analogy with the VSM \cite{Morchio88} and the CSM \cite{Carvalhaes97}, the proof of this assertion 
follows from the charge content of $ {\cal H} $ and the fact that 
some {\it topological}  charges get trivialized in the restriction 
from $ {\cal K}^{\,ext}_{_B} $ or $ {\cal K}_{_B} $ to $ {\cal H} $. 

To begin with, consider the massless dual 
field $ \tilde \Phi \equiv \{ \tilde \lambda^\prime,\tilde \chi_{_2} \}\,\in\,
\mbox{\boldmath $\Im$}^{ext}_{_B} $, defined by 
\\
$$ \partial _\mu \tilde \Phi (x) + \varepsilon_{\mu \nu} \partial
^\nu \Phi (x) = 0\,, \eqno (4.24) $$
\\
which is local with respect to itself but is non-local with respect 
to $ \Phi (x) $. The local charge 
operator $ \tilde {\cal Q}_{_{\tilde \Phi,R}} $
\\
$$ \tilde {\cal Q}_{_{{\tilde\Phi}}} = \lim_{_{R\rightarrow\infty}} 
\int\,d z^1\, \partial_{z^0} {\tilde\Phi }(z^0,z^1)\,f_{_R}(z^1)\,
\equiv \,\lim_{_{R \rightarrow \infty}} \tilde {\cal Q}_{_{{\tilde\Phi,R}}}\,,
\eqno(4.25)$$
\\ 
(with $ f_R(x^1)\,\in\,{\cal S}(\Re )$, $\lim_{R \rightarrow \infty} f_R(x) = 1$) is defined from 
the {\it topological} conserved 
current $ j_\mu (x) $ and obeys a local Gauss's 
law \cite{Strocchi74,Morchio88,Swieca76}
\\
$$ j_\mu (x) \equiv \partial _\mu \tilde \Phi (x) = \partial ^\nu
\varepsilon_{\nu \mu} \Phi (x) \equiv \partial ^\nu {\cal F}_{\nu
\mu}(x)\,. \eqno (4.26) $$
\\

The (local) charge 
operator $ \tilde {\cal Q}_{_{{\tilde \Phi}}} $ defines automorphisms
of the local  field algebras $ \mbox{\boldmath $\Im$}_{_B}^{ext}$, $ \mbox{\boldmath $\Im$}_{_B} $, $ \mbox{\boldmath $\Im$}
$. These automorphims are implementable in the corresponding 
spaces $ {\cal K}_{_B}^{ext} $, $ {\cal K}_{_B} $ and $ {\cal H} $ 
\cite{Morchio88,Morchio90}. Since $ \tilde \Phi (x) \,\in \,\mbox{\boldmath $\Im$}^{ext}_{_B} $, and
\\
$$ \lim_{R \rightarrow \infty}\,
[\,\tilde {\cal Q}_{_{{\tilde \Phi,R}}}\,,\,\mbox{\boldmath $\Im$}^{ext}_{_B}\,]\,
=\, [\,\tilde {\cal Q}_{_{{\tilde \Phi}}}\,,\,\mbox{\boldmath $\Im$}^{ext}_{_B}\,]\,\neq 0\,,
\eqno (4.27) $$
\\
this implies that the generators $ \tilde {\cal Q}_{_{{\tilde \Phi}}} $ do not
vanish on $ {\cal K}^{^{ext}}_{_B} $:
\\
$$ \tilde {\cal Q}_{_{{\tilde \Phi}}}\,{\cal K}^{ext}_{_B} \neq 0 \,.\eqno
(4.28) $$
\\

Since the intrinsic field algebra $ \mbox{\boldmath $\Im$} $ contains functions of the field 
operators $ \Phi (x) $, carried by $ \psi (x) $, and only derivatives of the
fields $ \tilde \Phi (x) $, through $ {\cal A}^L_\mu (x) $ and $ L_\mu(x) $, then 
\\
$$ \langle \tilde {\cal Q}_{_{\tilde\Phi}} \mbox{\boldmath{$\Psi_o$}} \, , \, 
\overline{\mbox{\boldmath $\Im$} \mbox{\boldmath{$\Psi_o$}}} \rangle = 0\,. \eqno(4.29)$$
\\ 
Hence we get 
\\
$$ \lim_{R \rightarrow \infty} 
\langle \tilde {\cal Q}_{_{ \tilde\Phi , R }} 
\mbox{\boldmath $\Im$} \mbox{\boldmath{$\Psi_o$}} \, , \, \overline{\mbox{\boldmath $\Im$} \mbox{\boldmath{$\Psi_o$}}} \rangle = $$

$$\lim_{R \rightarrow \infty} \langle [ \tilde {\cal Q}_{_{ \tilde\Phi , R }} 
\, , \, \mbox{\boldmath $\Im$} ] \mbox{\boldmath{$\Psi_o$}} \, , \, {\cal H} \rangle +  
\lim_{R \rightarrow \infty} \langle \mbox{\boldmath $\Im$} \tilde {\cal Q}_{_{ \tilde\Phi , R }} 
\mbox{\boldmath{$\Psi_o$}} \, , \, {\cal H} \rangle = 0.\eqno(4.30)$$
\\ 
In this way, for the set of local states $ {\cal D}_o \equiv \mbox{\boldmath $\Im$} \mbox{\boldmath{$\Psi_o$}} $,
we obtain the weak limit
\\
$$ \tilde {\cal Q}_{_{ \tilde\Phi }} {\cal D}_o = w - \lim_{R
\rightarrow \infty}  \tilde {\cal Q}_{_{ \tilde\Phi , R }} {\cal D}_o = 0\,,
\eqno(4.31)$$
\\ 
implying that the generators $ \tilde {\cal Q}_{\tilde \Phi} $ vanish
on $ {\cal H} $:
\\
$$  \tilde {\cal Q}_{_{ \tilde\Phi }} {\cal H} = 0\,.\eqno (4.32)$$
\\
This means that the charge $ \tilde {\cal Q}_{_{\tilde \Phi}} $ is
trivialized in the restriction from $ {\cal K}^{ext}_{_B} $ to $
{\cal H} $, i. e.,
\\
$$ \tilde {\cal Q}_{_{{\tilde \Phi}}}\, {\cal K}_{_B}^{ext} \neq 0\,\,\,,\,\,\,
\tilde {\cal Q}_{_{{\tilde \Phi}}}\, {\cal K}_{_B}\,
 =\, \tilde {\cal Q}_{_{{\tilde \Phi}}}\, {\cal H} = 0 \,. \eqno (4.33) $$
\\
The closure of local states associated to the field algebra $ \mbox{\boldmath $\Im$}
$ intrinsic to the model does not allow the introduction of operators
which are charged under $ \tilde {\cal Q}_{_{\tilde \lambda^\prime}}
$ and $ \tilde {\cal Q}_{_{\tilde \chi_{_2}}} $. In this way, the
states generated by $ \Omega_\ell (x) $, which are $ \tilde {\cal
Q}_{_{\tilde \chi_{_2}}} $- charged,
\\
$$ \lim_{_{R\rightarrow \infty}} [ \tilde {\cal Q}_{_{ \tilde\chi_{_2} , R }} \, , 
\, \Omega_{_\ell} ] = -i\sqrt{\pi} \Omega_{_\ell},\eqno(4.34)$$
\\ 
do not exist as solution of the subsidiary Gupta-Bleuler condition 
in ${\cal H}$:
\\
$$ \langle \Omega_{_\ell} \mbox{\boldmath{$\Psi_o$}} \, , \, {\cal H} \rangle =
\frac{i}{\sqrt{\pi}} \lim_{R \rightarrow \infty}
 \langle [ \tilde {\cal Q}_{_{ \tilde\chi_{_2},R }} \, 
, \, \Omega_{_\ell} ] \mbox{\boldmath{$\Psi_o$}} \, , {\cal H} \rangle =  $$
\\
$$ - 
\frac{i}{\sqrt{\pi}}\,\lim_{R \rightarrow \infty} \left\{ \langle \Omega_{_\ell}
 \tilde {\cal Q}_{_{\tilde\chi_{_2},R}} \mbox{\boldmath{$\Psi_o$}} \, ,  \, 
 {\cal H} \rangle - \langle \Omega_{_{\ell}} \mbox{\boldmath{$\Psi_o$}} \, , 
\, \tilde {\cal Q}_{_{ \tilde\chi_{_2},R }}  {\cal H} \rangle \right\} = 0\,. 
\eqno(4.35)$$
\\

Note that for $ ag_{_+}^2 - g_{_-}^2 \neq g_{_-}g_{_+} $ ( $ ag_{_+}^2 -
g_{_-}^2 \neq - g_{_-}g_{_+} $ ) the trivialization of the chiral 
charge $ {\cal Q}^5_L $ (associated with the longitudinal chiral
current $ L^5_\mu = \epsilon_{\mu \nu} L^\nu $) in the restriction
from $ {\cal K}^{ext}_{_B} $ to $ {\cal H} $,
\\
$$ {\cal Q}^5_L \,{\cal K}^{ext}_{_B} \neq 0\,\,,\,\,{\cal
Q}^5_L\,{\cal H} = 0\,,\eqno (4.36)$$
\\
implies that the closure of the local states associated to the
intrinsic field algebra $ \mbox{\boldmath $\Im$} $ does not allow the introduction of
operators which are charged under $ {\cal Q}^5_L $. However, for the
special case $ ag_{_+}^2 - g_{_-}^2 = g_{_-}g_{_+} $ ( $ ag_{_+}^2 -
g_{_-}^2 = - g_{_-}g_{_+} $ ) the operators $ \Lambda_{_\ell} (x) $,
$ \Gamma_{_r}(x) $, $ \sigma_{_\ell} (x) $, $\hat \sigma_{_r} (x) $, $ A(x) $, are neutral 
under $ {\cal Q}^5_L $ and the criterion based on
the trivialization of the charge $ {\cal Q}^5_L $ is insufficient to
decide about the existence of these operators 
as elements in $ {\cal H} $.

Although in the anomalous model the massless gauge field 
piece $ {\cal A}^L_\mu $ also belongs to the field 
subalgebra $ \mbox{\boldmath $\Im$}^\prime_{\chi_{_2},\lambda^\prime,\psi_o} $,  the
operator $ A (x) $ and the Wick exponential 
operator $ :\exp \{ - i A(x)\}: $, which also commute with the 
longitudinal current $ L_\mu $,  belong to the external Bose 
algebra $\mbox{\boldmath $\Im$}^{ext}_{_B}$, and cannot be defined as a solution of the 
subsidiary condition in $ {\cal H} $. These operators carry the 
charge $ \tilde {\cal Q}_{\tilde \chi_{_2}} $, which is
trivialized in the restriction from $ {\cal K}^{ext}_{_B} $ to $
{\cal H} $. On the other hand, the bilocal operator $ \Big (A (x) - A
(y)\Big ) $, as well as the Wick exponential 
operator $ : \exp \{ i e \int_x^y {\cal A}^L_\mu(\xi) d\xi^\mu \} : $,
used in the definition of the bilocal operators \footnote{\it See
discussion in subsection C below.},  can be defined as 
elements in $ {\cal H} $, since the line integral $ \int_x^y {\cal
A}^L_\mu(\xi) d\xi^\mu $ is  neutral with respect to both 
charges $ {\cal Q}_{_{\tilde \chi_{_2}}} $ and $ {\cal Q}_{_{
\tilde \lambda^\prime}} $ (as well as under $ {\cal Q}^5_L $). These operators
depend on $ \lambda^\prime (x) $ and thus 
are $ \tilde {\cal Q}_{_{\tilde \lambda^\prime}} $-neutral. The
neutrality under the charge  $ \tilde {\cal Q}_{_{\tilde \chi_{_2}}}
$ is due to the dependence on the neutral dipole
ghost configurations like $ \Bigl ( \tilde \chi_{_2} (x) - \tilde
\chi_{_2} (y) \Bigr ) $. This also ensures that the Wick exponential
of the line integral $ \int_x^y $ over the gauge field $ {\cal A}_\mu
$ can be used in the point-splitting limiting procedures to compute
composite operators belonging to the field algebra $ \mbox{\boldmath $\Im$} $, since this does 
not lead to the introduction of the field $\tilde
\chi_{_2}$, but only of its derivatives: 
\\
$$ \lim_{\epsilon \rightarrow 0}\,e_{_\ell}
\int_x^{x + \varepsilon}  {\cal A}^L_\mu(\xi) d \xi^\mu \propto
\varepsilon^\mu \partial_\mu \Bigl (\,\tilde \chi_{_2} (x) - \chi_{_2}
(x) + e_{_\ell} \lambda^\prime(x) \Bigr )\,.\eqno (4.37) $$
\\

Therefore, the operator $ \Omega_{_\ell}(x) $, which violates cluster 
property, cannot be defined in $ {\cal H} $. The
existence or not of operators like $ \sigma_{_\ell} $ in $ {\cal H} $
crucially depends on whether one uses or not the representation of the 
redundant field algebra $\mbox{\boldmath $\Im$}^{\, ext}_{_B}$, which contains
additional degrees of freedom not intrinsic to the model. The physical 
Hilbert space ${\cal H}$ does not 
contain  infinitely delocalized condensed states 
like $ \sigma{_\ell}^n\,\mbox{\boldmath{$\Psi_o$}} $, which carry the left free-fermion 
charge and chirality. The 
state $ \sigma^\ast_{_\ell} \mbox{\boldmath{$\Psi_o$}}$ does not belong to the
representation ${\cal H}$ of the intrinsic field algebra $\mbox{\boldmath $\Im$}$ and exists 
only in the redundant space ${\cal K}_{_B}^{\,ext}$. Consequently, the 
violation of the cluster decomposition and the need 
of a chiral $\Theta_\ell$-vacuum parametrizaton for the GCSM
defined with $ag_{_+}^2-g_{_-}^2=g_{_-}g_{_+}$, as suggested by the 
two-point Wightman functions of the 
field operator $ \Omega_\ell $, is deceptive. This is a consequence of 
an improper factorization of the completion of states and 
cannot be considered as a structural property of the GCSM.

In the chiral case, $ g_+ = g_- $ ($a=2$), although the field
component $ \lambda^\prime_\ell $ commutes with the longitudinal
current, the Wick exponential $ :e^{\,i g \lambda^\prime_\ell (x)}: $ cannot 
be defined in $ {\cal H} $ since it carries the charge $ \tilde {\cal
Q}_{_{\tilde \lambda^\prime}} $ which is trivialized in the
restriction from $ {\cal K}^{ext}_{_B} $ to $ {\cal H} $. Hence 
we cannot define the chiral density 
operator (4.17) as solution of the subsidiary condition in $ {\cal H}
$ \cite{Carvalhaes97}. 

In contrast to the vector model, in the 
anomalous case the bosonization scheme introduces a broader class
of operators belonging to $ \mbox{\boldmath $\Im$}_{_B} $, $ \mbox{\boldmath $\Im$}^{^{ext}}_{_B} $ which 
are not elements intrinsic to the model, that satisfy the subsidiary
condition. Consequently some care must be taken in order to construct
the Hilbert space $ {\cal H} $ representing the intrinsic field
algebra that defines the model. In particular, in the 
branch $ ag_+^2 - g_-^2 = g_+g_- $, one cannot 
define on $ {\cal H} $ the 
operators $ \hat\sigma_{_r}(x) $, $ \sigma_{_\ell}(x) $, $
\Gamma_{_r}(x) $, $\Lambda _{_\ell}(x) $, $ A (x) $, but only their
functions and derivatives appearing in $ \psi_{_\alpha} (x) $, $ W (x) $, $ {\cal
A}_\mu (x) $, etc.

In conclusion, a peculiar feature of the anomalous generalized 
chiral model which differs from
the vector case is the fact that the cluster decomposition property
is not violated for Wightman functions that are representations of
the intrinsic field 
algebra $\mbox{\boldmath $\Im$} = {\bf \mbox{\LARGE{\boldmath $\wp$}}} (\{\bar \psi,\psi,{\cal A}_\mu\})$, provided the
intrinsic field algebra alone is considered without any reduction of the
completion of states.

\section{Bilocals and the Subalgebra of Vector Schwinger Model}

In this section we shall make an analysis of the question
referring to adequacy of the
interpretation of the GCSM as an interpolating theory between chiral and pure
vector Schwinger models \cite{Basseto94,Alvaro94,Naon90}.

In order to try to ``reach'' at the quantum level the VSM by starting from the generalized
chiral model, we can proceed by two 
alternative limiting procedures. The first one is to 
set $ g_- = 0 $ ( $ e_{_\ell} = e_{_r} \equiv e $, $ m^2 =
(e_{_\ell}^2/ \pi )(a + 1) $ ) and then take 
the limit $ a \rightarrow 0^+ $. Alternatively,  we can consider the 
branch $ a g^2_+ - g^2_- = g_- g_+ $, for 
which $ b = 1 $, $ m^2 =   e^2_{_\ell}/ \pi $, and then 
set $ g_- = 0 $, which implies simultaneously that $ a = 0 $.  However, these 
limiting procedures are not defined for the whole
set of Wightman functions that define the GCSM, since in general, they carry 
the contributions coming from the field $ \lambda^\prime $ and, as (2.7)
shows, its Wightman functions diverge in this limit and thus are 
ill-defined (see for example the two-point function (2.16)). As a matter of 
fact this
limit when performed in the quantum theory is meaningless since the 
quantization of the model has been carried out assuming that $ a
g_+^2 - g_-^2 \neq 0 $, which ensures a non trivial dynamical nature for 
the field $ \lambda^\prime $. Analogous situation occurs in the CSM 
for which we have
different algebraic constraint structures leading to distinct quantum
theories corrersponding to the cases $ a = 0 $, $ a = 1 $ and $ a > 1 
$ \cite{RRG86}. The singular character of this kind of ``limit''  means 
that we are dealing with two inequivalent classes 
of distinctly constrained theories which  can
only be connected through an infinite weighted gauge 
transformation. 

Performing the 
limit $ g_- \rightarrow 0 $ ($ b = 0 $), $ a \rightarrow 0^+ $, in
the operator solution (2.9), we get, besides the singular 
dependence on the field $
\lambda^\prime $, the Lowenstein-Swieca solution of the vector SM \footnote{\it
The notation of Ref. \cite{Lowenstein71} is : 
$\{\chi_{_1},\chi_{_2},\phi^\prime,\lambda^\prime\} \rightarrow \{\tilde
\Sigma,- \tilde \eta,\phi,0\} $.} \cite{Lowenstein71}
\\
$$ \psi_\alpha(x) = :\psi_\alpha^{^o}(x)\,e^{\,i \sqrt \pi\,
\gamma^5_{\alpha\alpha}\,\Bigl ( \chi_{_1}(x) - \chi_{_2}(x) \Bigr )}:\,:e^{
\,i \lambda^\prime(x)}:\,,\eqno (5.1a)$$
\\
$${\cal A}^\mu(x) = - \frac{\sqrt \pi}{e}\,
\tilde{\partial}^\mu \Bigl (\chi_{_2}(x)-\chi_{_1}(x) \Bigr ) + 
\partial^\mu\lambda^\prime(x)\,,\eqno (5.1b)$$
\\
$$ L^\mu(x) = - \frac{1}{\sqrt{\pi}}\,\tilde{\partial}^\mu\,
\Bigl ( \phi^\prime (x) + \chi_{_2}(x) \Bigr )\,.\eqno (5.1c) $$
\\
and the following local algebraic restrictions
\\
$$ \Bigl [ \psi (x),L_\mu(y) \Bigr ] \neq 0\,\,,\,\,\Bigl [ {\cal
A}_\mu(x),L_\nu(y) \Bigr ] \neq 0\,.\eqno (5.2) $$
\\
From the above commutation relations it can be observed that the 
algebraic constraint structure of the model was
changed in order to accommodate the appearance of a genuine gauge invariant 
subalgebra $ \mbox{\boldmath $\Im$}_{phys}\,\subset\,\mbox{\boldmath $\Im$} $. In this ``limit'' the field
$\lambda^\prime$ loses its algebraic (local operator) content and gives
rise to divergent Wightman functions.  The attempt to gauge away the
contributions coming from the field $ \lambda^\prime $ can only be 
implemented by an ill-defined singular operator gauge transformation.

Nevertheless,  one can obtain a field subalgebra
that does not contain the field $ \lambda^\prime $ as element. This
field subalgebra is constructed from the fundamental set of field
operators $ \{\bar \psi,\psi,{\cal A}_\mu\} $ and the exponential 
operator of the line integral over the gauge field. However, the construction 
of this field subalgebra involves delicate mathematical aspects
and some care must be exercised.

In the branch $ ag_+^2 - g_-^2 = g_- g_+ $,  we can consider a gauge
invariant field subalgebra $ \mbox{\boldmath $\Im$}^{^{g\i}} \,\subset \,\mbox{\boldmath $\Im$} $, which does
not contain the field $ \lambda^\prime (x) $ as element. Consider  the
formally ``gauge invariant'' bilocal operators \cite{Lowenstein71}
\\
$$ D_{\alpha \alpha}(x,y) = : \psi_\alpha (x)\,e^{\,
i e_\alpha \int_x^y {\cal A}_\mu(\xi) d \xi^\mu }\,\psi_\alpha^\ast (y) :
\,,\eqno (5.3) $$
\\
which are given by
\\
$$ D_{\ell \ell}(x,y) = \Big ( \frac{\mu_o}{2 \pi} \Big )\, : \,e^{\, i \sqrt \pi \Bigl \{ \chi_{_1} (x) 
 + \int_x^y\, \epsilon_{\mu \nu} \partial^\nu \chi_{_1}(\xi) d \xi^\mu -
\chi_{_1} (y) \Bigr \}}\,:\, \sigma_\ell(x) \sigma^\ast_\ell(y)\,,
\eqno (5.4a) $$
\\
$$ D_{r r}(x,y) =  \Big ( \frac{\mu_o}{2 \pi} \Big )\,: \,e^{\, 
i \frac{e_r}{e_\ell}\,\sqrt \pi \Bigl \{ \chi_{_1} (x) 
 + \int_x^y\, \epsilon_{\mu \nu} \partial^\nu \chi_{_1}(\xi) d \xi^\mu -
\chi_{_1} (y) \Bigr \}}\,:\,\times
$$
\\
$$ \times\,:e^{\,2 i \sqrt \pi \Bigl \{
\phi^\prime_{_r} (x) + \frac{e_r}{e_\ell} \chi_{_{2r}} (x) \Bigr \}}:
\,:e^{\,- 2 i \sqrt \pi \Bigl \{
\phi^\prime_{_r} (y) + \frac{e_r}{e_\ell} \chi_{_{2r}} (y) \Bigr
\}}:\,. \eqno (5.4b) $$
\\
The operator $ D_{\ell \ell} $ corresponds to the Lowenstein-Swieca 
bilocal operator of the VSM \cite{Lowenstein71}. 

Although the 
operator $ \sigma_\ell (x) $ cannot be defined by itself
in  $ {\cal H} $, in virtude of
\\
$$ \Big [ \tilde {\cal Q}_{\tilde \chi_{_2}} , \sigma_\ell (x) \Big ] = - 2
\sqrt \pi \sigma_\ell (x) \eqno (5.5)$$
\\
the ``composite'' spurious (neutral) dipole 
operator $ \sigma^\ast_\ell(x) \sigma_\ell (y) $ is defined as an
element of $ {\cal H} $ and leads to constant vacuum expectation value
\\
$$ \langle \mbox{\boldmath{$\Psi_o$}}, \sigma^\ast_\ell(x) \sigma_\ell(y) \mbox{\boldmath{$\Psi_o$}} \rangle
= 1 \,.\eqno (5.6) $$
\\
The state $ (\sigma^\ast_\ell \sigma_\ell) \mbox{\boldmath{$\Psi_o$}} $ is translationally
invariant in $ {\cal H}$. The position independence of this state can be 
seen by computing the general Wightman
funcions involving the operator $ \sigma^\ast_\ell(x) \sigma_\ell(y)
$ and all operators belonging to the local field algebra $ \mbox{\boldmath $\Im$} $. Thus, for 
any operator $ {\cal O}(f_{z}) = \int \,{\cal O}(z)
f(z)\,d^2 z\,\in\,\mbox{\boldmath $\Im$} $ of polynomials in the
smeared fields $\{\bar\psi,\psi,{\cal A}_\mu\}$, the position
independence of the operator $\sigma^\ast_\ell \sigma_\ell $ can be 
expressed in the weak form as 
\\
$$ \langle\,\mbox{\boldmath{$\Psi_o$}}, \sigma^\ast_\ell(x)\sigma_\ell(y) \,
{\cal O}(f_{z_1},\cdots,f_{z_n})\,\mbox{\boldmath{$\Psi_o$}}\,\rangle =  $$
\\
$$ = {\cal W} (z_1,\cdots,z_n) \equiv 
\langle\,\mbox{\boldmath{$\Psi_o$}},{\cal O}(f_{z_1},\cdots,f_{z_n})\,\mbox{\boldmath{$\Psi_o$}}\,\rangle
\,\,\,,\,\,\,\forall\,{\cal O}(f) \in \mbox{\boldmath $\Im$}\,, \eqno (5.7)$$
\\
where $ {\cal W} (z_1,\cdots,z_n) $ is a distribution (Wightman function) independent 
of the space-time coordinates $(x,y)$. The spurious 
operator $ \sigma^\ast_\ell \sigma_\ell $  does
not carry any charge selection rule, and since it commutes with all operators
belonging to the field algebra $ \mbox{\boldmath $\Im$} $, it is reduced to the 
identity operator in $ {\cal H} $. In this way, in spite of the existence  
of the constant operator $ \sigma_\ell^\ast \sigma_\ell $ in the field algebra,  the cluster 
decomposition property is not violated.

Since the gauge invariant field subalgebra $ \mbox{\boldmath $\Im$}^{^{g\i}} $ does not
contain the field $ \lambda ^\prime(x) $ as element, the ``vector
limit'' $ e_r \rightarrow e_\ell = e $, can be performed for
operators belonging to this field subalgebra. For the right component
$ D_{r r} $, given by (5.4b), in this limit we obtain the bilocal
operator of the Lowenstein-Swieca solution of the VSM \cite{Lowenstein71}
\\
$$ D_{r r}(x,y) =  \Big ( \frac{\mu_o}{2 \pi} \Big )\,: \,e^{\, 
i \,\sqrt \pi \Bigl \{ \chi_{_1} (x) 
 + \int_x^y\, \epsilon_{\mu \nu} \partial^\nu \chi_{_1}(\xi) d \xi^\mu -
\chi_{_1} (y) \Bigr \}}\,:\,\sigma_{_r}(x) 
\sigma_{_r}^\ast (y)\,,\eqno (5.8)$$
\\
where $  \sigma_{_r} $ is the spurion operator that appears in the 
Lowenstein-Swieca solution of the VSM \cite{Lowenstein71} 
\\
$$ \sigma_{_r}(x) = : e^{\,2i\sqrt \pi \Big [ \phi_{_r}^\prime
(x) + \chi_{_{2r}} \Big ]}:\,, \eqno (5.9) $$
\\
and carries the right free-fermion chirality. However, this limit cannot 
be performed for the whole set of Wightman
functions that define the generalized anomalous chiral model.

Another strategy is to consider a ``gauge noninvariant'' field subalgebra 
$ \mbox{\boldmath $\Im$}^{^{gn\i}} \subset \mbox{\boldmath $\Im$} $, that includes the field $
\lambda^\prime $ as element and which in the vector limit
smoothly reduces to the gauge invariant bilocals of the VSM. Consider
for instance the bilocal operator that is gauge invariant only in this 
limit:
\\
$$ \tilde D_{r r}(x,y) = : \psi_{_r} (x)\,e^{\,
i e_\ell \int_x^y {\cal A}_\mu(\xi) d \xi^\mu }\,\psi_{_r}^\ast (y) :
= $$
\\
$$ = \Big ( \frac{\mu_o}{2 \pi} \Big )\,: \,e^{\, 
i \,\sqrt \pi \Bigl \{ \chi_{_1} (x) 
 + \int_x^y\, \epsilon_{\mu \nu} \partial^\nu \chi_{_1}(\xi) d \xi^\mu -
\chi_{_1} (y) \Bigr \}}\,\times $$
\\
$$ \times :\, \sigma_{_r}(x)
 \sigma_{_r}^\ast(y)\,:e^{\,i \Bigl \{\,(e_r - e_\ell) \lambda^\prime (x) -
(e_r - e_\ell) \lambda^\prime (y)\,\Bigr \}}: \,. \eqno (5.10) $$
\\
After ``taking the limit''  to  the  vetor model, the operator
(2.58) becomes gauge invariant and also maps onto the bilocal
fields of the VSM. Since the dependence on the field $\lambda^\prime$
is suppressed, one can ``take'' the limit $ a \rightarrow 0 $ in the
field subalgebra $\mbox{\boldmath $\Im$}^{^{gn\i}}$.

Although in this limit a field subalgebra of
the GCSM maps smoothly into the algebra of the gauge invariant operators of 
the VSM, this limit is not fully defined for the general Wightman functions of
the original anomalous  model and thus is ill-defined for the field
algebra. Owing to the fact that the so-called vector
limit cannot be performed on the Hilbert space that provides a
representation of the intrinsic
local field algebra, rigorously we cannot consider the GCSM as an 
interpolating quantum 
theory
between pure vector and chiral Schwinger models. As a matter of fact, in 
order to accommodate the intrinsic field algebra to a 
new constraint class defining a genuine gauge invariant
subalgebra, a singular operator gauge transformation must be involved to gauge
away the degrees of freedom carried by the field $ \lambda^\prime $. A similar 
situation occurs in the zero-mass limit of the Thirring-Wess model
\cite{Lowenstein71,Rothe77}.

\section{Extended Gauge Invariant Formulation}

In this section we shall give briefly the guidelines for the
generalization of the construction of the $GI$ formulation of the chiral model
given in Refs. \cite{Boyanovsky88,Carvalhaes97} for the GCSM. This
generalizes and completes the presentation of Ref.\cite{Boyanovsky88}.

The so-called $GI$ formulation of an anomalous
gauge theory is constructed introducing extra degrees of freedom into
the theory by ``adding'' to the original $GNI$ Lagrangian a Wess-Zumino
($WZ$) term \cite{Faddeev86,Wess71}. Within the
path-integral \cite{Babelon86,Girotti89} and operator \cite{Belvedere95}
approaches, the $GI$ formulation is constructed by
enlarging the intrinsic field algebra $ \mbox{\boldmath $\Im$} \equiv \mbox{\boldmath $\Im$}_{_{GNI}} $ through an
operator-valued gauge transformation on the $GNI$ formulation of the model:
\\
$$ ^\theta\!\psi(x) = e^{\,\frac{i}{2}\left[e_{_r}(1+\gamma^5)+e_{_\ell}
(1-\gamma^5)\right]\theta(x)} \,\psi(x)\,, \eqno(6.1a)$$

$$ ^\theta\!{\cal A}_\mu(x) = {\cal A}_\mu(x) + \partial_\mu \theta(x)\,.
\eqno (6.1b)$$
\\
The gauge-transformed Lagrangian density $ ^\theta\!{\cal L} \equiv
{\cal L}_{_{GI}} $ is given by
\\
$$ {\cal L}_{_{GI}}\{^\theta\!\bar \psi, ^\theta\!\psi, ^\theta\!{\cal A}_\mu\} = 
{\cal L}_{_{GNI}}\{\bar \psi, \psi, {\cal A}_\mu\} + {\cal
L}_{_{WZ}}\{{\cal A}_\mu, \theta\}\,, \eqno (6.2)$$
\\
where $ {\cal L}_{_{WZ}} $ is the WZ Lagrangian density
\\
$$ {\cal L}_{_{WZ}} = \frac{1}{2}\,\frac{g_-^2}{\pi}\,
\Biggl ( a \frac{g_+^2}{g_-^2} - 1\Biggr ) (\partial _\mu \theta ) ^2 + 
\frac{1}{\sqrt \pi}\,\frac{g_-}{\sqrt \pi}\, \Biggl
[ \Biggl ( a \frac{g_+^2}{g_-} - g_- \Biggr ) \partial _\mu \theta -
g_+ \tilde \partial _\mu \theta \Biggr ] {\cal A}^{\,\mu}\,.\eqno (6.3) $$
\\
The resulting ``embedded'' theory exhibits invariance under the {\it
extended} local gauge transformations
\\
$$\psi(x)\rightarrow e^{\,\frac{i}{2} \left[e_{_r}(1+\gamma^5)+e_{_\ell}
(1-\gamma^5)\right]\Lambda(x)}\psi(x)\,, \eqno(6.4a)$$

$${\cal A}_\mu(x)\rightarrow {\cal A}_\mu(x) + \partial_\mu\Lambda(x)\,, 
\eqno (6.4b)$$

$$ \theta (x) \rightarrow \theta (x) - \Lambda (x)\,. \eqno (6.4c) $$
\\
The set of gauge-transformed field 
operators $\{ ^\theta\!\bar\psi,^\theta\!\psi,^\theta\!\!{\cal A}_\mu \}$ is
invariant (by construction) under the extended local gauge
transformations (3.4) and generates the $GI$ field algebra $^\theta\mbox{\boldmath $\Im$}
\equiv \mbox{\boldmath $\Im$}_{_{GI}}$. We  shall denote 
by $\,^\theta{\cal H}\,\equiv\,{\cal H}_{_{_{GI}}}\,\doteq\,^\theta \mbox{\boldmath $\Im$}\,
\mbox{\boldmath{$\Psi_o$}} $, the state space 
on which the field algebra generated from the set of
gauge-transformed field operators $ \{\,^\theta\!\bar \psi,\,
^\theta\!\psi,\,^\theta\!\!{\cal A}_\mu\,\} $ is represented.

Proceeding along the same lines as those of Refs. \cite{Boyanovsky88,Carvalhaes97}, we consider the enlargement of the 
Bose field algebra $ \mbox{\boldmath $\Im$} ^{^B}_{_{_{GNI}}} $ by the introduction of the WZ 
field through the operator gauge transformation (6.1). This
corresponds to shifting the field $ \lambda^\prime $ in the bosonized
expressions for the field operators (2.9) by the {\it extended} gauge invariant
combination $ \lambda^\prime + \theta $. Following Refs.
\cite{Boyanovsky88,Carvalhaes97}, we define a new 
field $ \xi^\prime \doteq \lambda^\prime + \theta $ such that, in the 
effective bosonized theory, the field $\lambda^\prime$ is
replaced by  the new field $ \xi^\prime $, with the fields satisfying the 
algebraic constraints
\\
$$ [\,\lambda^\prime (x)\,,\,\xi ^\prime (y)\,]\,=\,0\,\,,\eqno (6.5a) $$
\\
$$ [\,\theta (x)\,,\,\xi ^\prime (y)\,]\,=\,-\,
[\,\theta (x)\,,\,\lambda ^\prime (y)\,]\,=\,[\,\xi ^\prime (x)\,,\,
\xi ^\prime (y)\,]\,=\,[\,\lambda ^\prime (x)\,,\,\lambda ^\prime
(y)\,]\,\,. \eqno (6.5b) $$
\\
As shown in Ref. \cite{Carvalhaes97}, the shift $\xi^\prime$, together with 
the above commutation relations leads to a gauge invariant algebra and 
ensures that no additional physical degree of freedom is introduced into the 
theory. In this way, and analogously to what happens in the chiral
case, the 
field algebra $ \mbox{\boldmath $\Im$}_{_{_{GNI}}}^{^B} $ of the $GNI$ formulation
generated from the building 
blocks $ \{\,\phi^{\prime}, \chi_{_2},\chi_{_1}, \lambda^{\prime}\,\} $ is 
replaced in the $GI$ formulation by the 
algebra $ \,\mbox{\boldmath $\Im$}_{_{_{GI}}}^{^B} $ generated from the set of 
fields $ \{\,\phi^{\prime}, \chi_{_2},\chi_{_1}, \xi^{\prime}\,\} $. Using 
the algebraic constraints (6.5), we can
display the following implementability  conditions 
\\
$${\cal Q}_{_{_{\lambda^\prime}}}{\cal H}^{^B}_{_{_{GNI}}}\,\neq\,0\,\,\,,
\,\,\,{\cal Q}_{_{_{\lambda^\prime}}}{\cal H}_{_{_{GNI}}}\,\neq\,0\,\,\,, $$
$${\cal Q}_{_{_{\lambda^\prime}}}{\cal H}^{^B}_{_{_{GI}}}\,=\,0\,\,\,,
\,\,\,{\cal Q}_{_{_{\lambda^\prime}}}{\cal
H}_{_{_{GI}}}\,=\,0\,\,\,;\eqno (6.6a)$$
\\
$${\cal Q}_{_{_{\xi^\prime}}}{\cal H}^{^B}_{_{_{GNI}}}\,=\,0\,\,\,,
\,\,\,{\cal Q}_{_{_{\xi^\prime}}}{\cal H}_{_{_{GNI}}}\,=\,0\,\,\,,$$
$${\cal Q}_{_{_{\xi^\prime}}}{\cal H}^{^B}_{_{_{GI}}}\,\neq\,0\,\,\,,
\,\,\,{\cal Q}_{_{_{\xi^\prime}}}{\cal H}_{_{_{GI}}}\,\neq\,0\,\,\,;
\eqno (6.6b)$$
\\
and 
since $ \theta\,=\,\xi^\prime\,-\,\lambda^\prime $, we get
\\
$${\cal Q}_{_{_{\theta}}}{\cal H}^{^B}_{_{_{GNI}}}\,\neq\,0\,\,\,,
\,\,\,{\cal Q}_{_{_{\theta}}}{\cal H}_{_{_{GNI}}}\,\neq\,0\,\,\,,$$
$${\cal Q}_{_{_{\theta}}}{\cal H}^{^B}_{_{_{GI}}}\,\neq\,0\,\,\,,
\,\,\,{\cal Q}_{_{_{\theta}}}{\cal H}_{_{_{GI}}}\,\neq\,0\,\,\,.\eqno (6.6c) $$
\\
The latter
condition implies that the Wick exponential and derivatives of the WZ
field $ \theta $ can be defined on both Hilbert 
spaces $ {\cal H}_{_{_{GNI}}} $ and $ {\cal H}_{_{_{GI}}} $, as
implied by the operator-valued gauge transformation (6.4) which
enables the relationship between the isomorphic $GNI$ and $GI$
formulations. This is understandable from the observation that in the
anomalous model the field operators $ \{ \psi, {\cal A}_\mu \} $
acquire the status of physical observables, and any acceptable
transformation on them must be innocuous as concerns the physical
content of the theory.

The formal expressions 
for the operators $ \{\,\psi_{_{_{GI}}},\,{\cal A}^\mu_{_{_{GI}}},\,
L^\mu_{_{_{GI}}}\,\} $ are the same as for $ \{\,\psi_{_{_{GNI}}},\,
{\cal A}^\mu_{_{_{GNI}}},\,L^\mu_{_{_{GNI}}}\,\} $, except for the
replacement of the 
field $ \lambda ^\prime $ by the field $ \xi ^\prime $. The 
set 
of field operators $ \{ \bar\psi _{_{_{GI}}},\,\psi _{_{_{GI}}},\,
{\cal A}^\mu _{_{_{GI}}} \} $ defines (through polynomials of these 
smeared fields, Wick ordering, point-splitting regularization
of polynomials, etc.) the $GI$ algebra $ \mbox{\boldmath $\Im$}_{_{_{GI}}} $ which is subject 
to the constraint
\\
$$ [\,{\cal O}\,,\,L^{\mu }_{_{_{GI}}}\,]\,=\,0\,\,\,,\,\forall\,{\cal O}\,\in
\,\mbox{\boldmath $\Im$}_{_{_{GI}}} \,\,\,.\eqno (6.7) $$
\\
Although the so-introduced WZ field has acquired dynamics, it is a redundant 
field in the sense that it does not change the algebraic structure of
the model and therefore does not change its physical content. As
shown in Ref. \cite{Carvalhaes97}, this
implies the isomorphism between the field algebras $ \mbox{\boldmath $\Im$}_{_{_{GNI}}}
$ and $ \mbox{\boldmath $\Im$}_{_{_{GI}}} $.

The isomorphism between the field algebra $ \mbox{\boldmath $\Im$}_{_{_{GNI}}} $ defining the $GNI$
formulation and the algebra $ \mbox{\boldmath $\Im$}_{_{_{GI}}} $ defining the $GI$
formulation of the CSM implies that the state 
space $ {\cal K}_{_{_{GNI}}} $, which provides a representation of the $GNI$ intrinsic
local field algebra $ \mbox{\boldmath $\Im$}_{_{_{GNI}}} $, is isomorphic to the state
space $ {\cal K}_{_{_{GI}}} $ on which the $GI$ local field algebra is 
represented; i.e.,
\\
$$ \langle\,{\bf \mbox{\LARGE{\boldmath $\wp$}}} \,\{\,\bar\psi_{_{_{GNI}}},\,\psi_{_{_{GNI}}},\,
{\cal A}^\nu_{_{_{GNI}}}\,\}\,\rangle\,\equiv\,
\langle\,{\bf \mbox{\LARGE{\boldmath $\wp$}}} \,\{\,\bar\psi_{_{_{GI}}},\,\psi_{_{_{GI}}},\,
{\cal A}^\nu_{_{_{GI}}}\,\}\,\rangle\,\equiv \langle\,
{\bf \mbox{\LARGE{\boldmath $\wp$}}} \,\{\,^\theta\!\bar\psi,\,^\theta\!\psi,\,^\theta\!\!{\cal A}^\nu\,\}\,
\rangle\,\,, \eqno (6.8) $$
\\
where {\Large $ {\bf \mbox{\LARGE{\boldmath $\wp$}}} $} is any polynomial in the intrinsic field
operators. 

The same analysis made in section II for the $GNI$ formulation
applies to its isomorphic $GI$ formulation. The conclusions referring
to cluster decomposition and the would-be interpolation between the
chiral and pure vector models are the same for the $GI$ formulation.

\section{Conclusions}

We have investigated the physical content of the generalized chiral 
Schwinger model by means of a careful analysis of its 
mathematical structure. The Hilbert 
space has been constructed as the representation of the intrinsic field algebra 
generated by the basic set of field operators whose Wightman functions define 
the model. The use of a redundant field algebra has
significant effects in the derivation of the fundamental physical properties 
of the model. In particular, we have made clear that in the second 
range (2.8b) of the Jackiw-Rajaraman parameter the model loses most of its 
physical meaning, and statements concerning the screening of electric 
charge, for example, are unjustifiable. Furthermore, we have argued 
that, strictly speaking, the vector Schwinger model cannot be viewed 
as a limit of the generalized chiral Schwinger model because the limit 
can only be properly defined for a local field subalgebra of the 
generalized model. We  have  displayed  in different  fashions the
construction of the bilocal  operators  that in the  vector  limit  
represent the
observable  content of  the Schwinger model  and  discussed  in  detail
whether   they  belong or  not to the algebra of observables of the generalized
chiral model. The gauge-invariant 
formulation of the generalized model has been constructed and, at the level 
of the respective field algebras, it was shown to be isomorphic to the 
gauge-noninvariant formulation.

We conclude by remarking that recent research \cite{Marino97} points 
towards the extension of the bosonization scheme to higher space-time
dimensions, and a foundational investigation using the general strategy 
of embedding the mathematical structures of the bosonization scheme
into the context of the principles of the general theory of quantized
fields may offer a valuable 
lesson in the study of the physical content of more realistic 
theories.

\underline{{\it Acknowledgments}}: The authors are grateful to 
Conselho Nacional de Desenvolvimento
Cient\'{\i}fico e Tecnol\'ogico (CNPq - Brasil) and
Coordena\c{c}\~ao de Aperfei\c{c}oamento de Pessoal de N\'{\i}vel
Superior (CAPES- Brasil) for partial financial support.

\end{document}